\documentclass[AMA,Times1COL]{WileyNJDv5}

\articletype{Article Type}%

\usepackage{caption}

\usepackage{pgfplots}
\usepackage{multirow}
\usepackage{listings}
\usepackage{pgfplots}
\pgfplotsset{compat=1.18}
\usepackage{amssymb,amsmath,amsthm,amsfonts}
\usepackage{calc}
\usepackage{graphicx}
\usepackage{subcaption}
\usepackage{indentfirst}
\usepackage{tikz}
\usepackage{url} 
\usepackage{circuitikz}
\usepackage{titlesec}
\usepackage{amsmath, amssymb, amsthm}
\usepackage{bm}
\usepackage{colortbl}   
\usepackage{xcolor}     
\usepackage{bm}
\usepackage{siunitx}
\usepackage{enumitem}
\usepackage{steinmetz}
\usepackage{multirow}
\usepackage{array}
\usepackage{tabularx}
\usepackage{listings}
\usepackage{mathtools}
\usepackage{cleveref}

\newcommand{\Th}{\widehat{\theta}}
\newcommand{\ph}{\widehat{p}}

\newcommand{\bq}{\mathbf{q}}

\DeclareGraphicsExtensions{.bmp,.png,.pdf,.jpg}
\usepackage{gensymb}
\usepackage{notoccite}
\usepackage{wrapfig}
\DeclareMathSymbol{\dr}{\mathalpha}{operators}{`d}

\received{Date Month Year}
\revised{Date Month Year}
\accepted{Date Month Year}
\journal{International Journal for Numerical Methods in Biomedical Engineering}
\volume{00}
\copyyear{2025}
\startpage{1}

\begin{document}

\title{A Physics-Informed Neural Network Approach to Multiphysics Continuum Modeling of Cancer Growth via Chemo-fluid Coupling}

\author[1]{Celia Taboada}

\author[1]{Pedro Navas}

\author[1]{Miguel Molinos}

\authormark{TABOADA \textsc{et al.}}
\titlemark{A Physics-Informed Neural Network Approach to Multiphysics Continuum Modeling of Cancer Growth via Chemo-fluid Coupling}

\address[1]{\orgdiv{Continuum Mechanics and Theory of Structures Dep. Civil Eng. School}, \orgname{Universidad Politécnica de Madrid}, \orgaddress{\state{Madrid}, \country{Spain}}}

\corres{Pedro Navas \email{pedro.navas@upm.es}}



\abstract[Abstract]{Tumor progression is an inherently multiphysical phenomenon in which interstitial fluid dynamics, biochemical transport, and cellular mechanics interact across multiple spatiotemporal scales. Classical mesh-based solvers, although accurate, impose prohibitive computational costs for the repeated evaluations demanded by inverse parameter identification and
future patient-specific predictive pipelines. In this work we introduce a
Physics-Informed Neural Network (PINN) framework for a tractable
\emph{chemo-fluidic} continuum model of tumor growth that couples an
advection-diffusion-reaction (ADR) equation for the tumor volume fraction
$\theta$ with a quasi-static Darcy pressure equation for the interstitial
fluid pressure $p$. By intentionally decoupling the solid-mechanical
equilibrium, we obtain a three-equation system whose gradient structure
is stable under automatic differentiation, enabling robust deep-learning
optimization. The network simultaneously learns both state variables
$(\theta, p)$ from physics constraints alone (forward problem) and
recovers hidden transport parameters from sparse, noisy synthetic
measurements (Data-Assimilation PINN, DA-PINN, inverse problem). We
verify the forward solver against a high-resolution finite-difference
(FD) reference, achieving a mean absolute error of $\bar{e}_{\Th} <
0.002$. For the inverse problem, starting from an initial permeability
estimate $K_{\mathrm{init}} = 0.08$ (a factor of $4\times$ above the
true value $K_{\mathrm{true}} = 0.02$), with only 5\,\% spatially sparse
observations corrupted by 5\,\% Gaussian noise, the DA-PINN recovers the
permeability with a relative error below 5\,\%. These results demonstrate
that physics-informed deep learning constitutes a viable, computationally
efficient route to multiphysics oncology modeling and lays the
mathematical groundwork for future integration into clinical data
assimilation pipelines.}


\jnlcitation{\cname{%
\author{Taboada C.},
\author{Navas P.},
\author{Molinos M.},}.
\ctitle{A Physics-Informed Neural Network Approach to Multiphysics Continuum Modeling of Cancer Growth via Chemo-fluid Coupling.} \cjournal{\it IJNMBE} \cvol{2025;00(00):1--16}.}

\maketitle



\section{Introduction}
\label{sec:intro}
Tumor progression is not exclusively a biological or genetic perturbation; it is fundamentally a coupled problem in continuum mechanics. As malignant cells proliferate within a confined tissue matrix, they exert solid stress on the surrounding parenchyma and simultaneously act as pressure sources for the interstitial fluid, driving a pathological elevation of the \emph{interstitial fluid pressure} (IFP) \cite{Jain2014}. This elevated IFP, which can reach values of 5--40\,mmHg in solid tumors compared to near-zero in healthy tissue, generates steep pressure gradients at the tumor periphery that give rise to convective interstitial flows \cite{Jain1988}. These advective fluxes act as a physical barrier to the inward perfusion of systemically administered chemotherapeutic agents, promoting drug exclusion and thereby contributing directly to therapeutic resistance \cite{Stylianopoulos2012, Chauhan2011}. Simultaneously, the outward convective current facilitates the escape of tumor cells into the peri-tumoral stroma, providing a fluid-mechanical pathway for local invasion \cite{Polacheck2011}. A rigorous theoretical description of these phenomena was established by Jain et al.\cite{Jain1988} within the framework of Starling forces and interstitial fluid mechanics, and further elaborated by \cite{Pozrikidis2003} through the application of porous-medium flow theory to biological tissues. \cite{Roose2003} provided the first systematic analysis of interstitial fluid pressure within poroelastic tumor models, quantifying the relationship between mechanical stress, tissue permeability, and IFP elevation. These foundational contributions established the theoretical consensus that any mechanistically faithful model of the tumor microenvironment (TME) must integrate cellular biology with interstitial fluid dynamics---a coupling that purely reaction-diffusion approaches cannot capture.

\begin{figure}[htbp]
\centering
\includegraphics[width=0.7\linewidth]{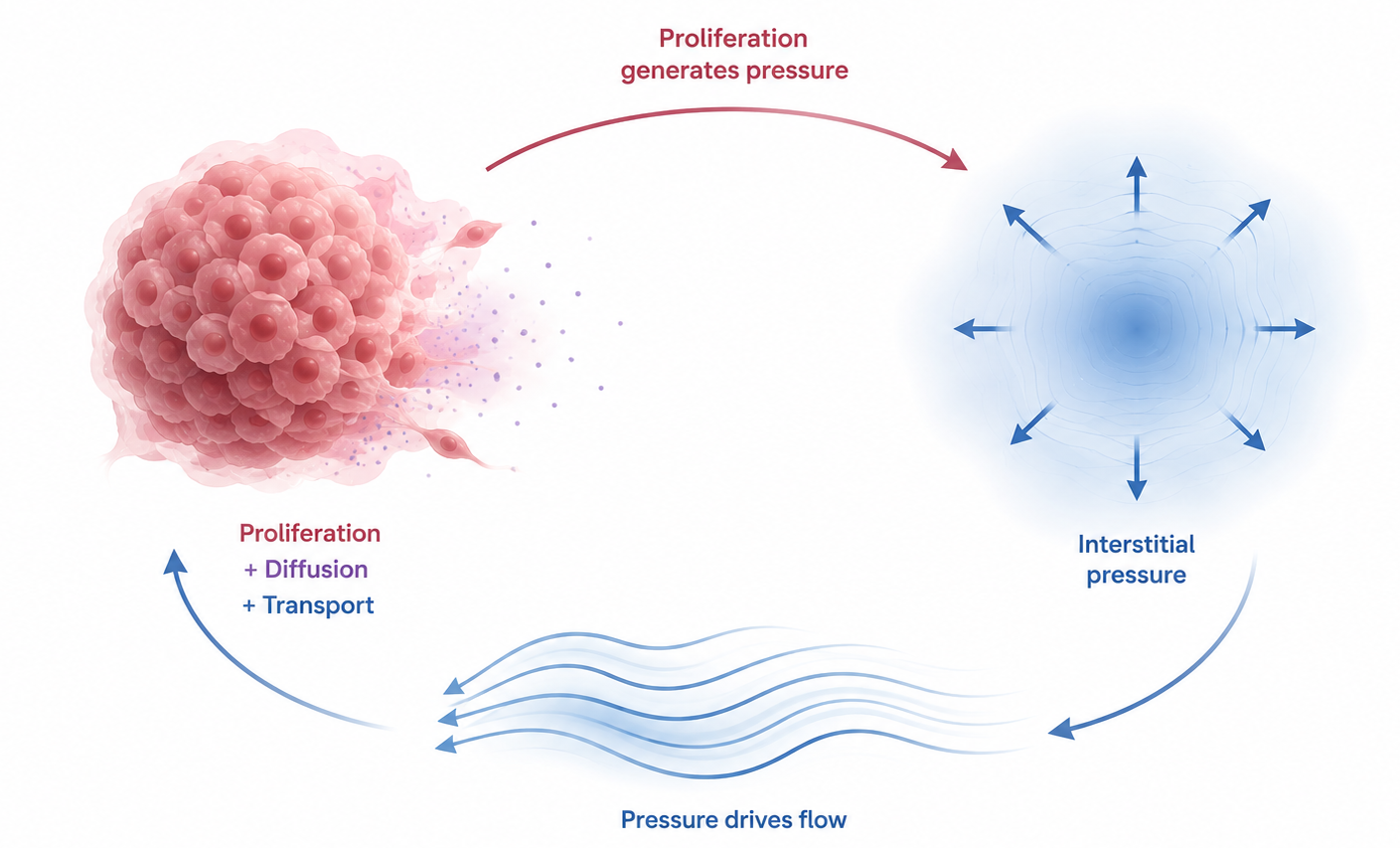}
\caption{Conceptual chemo-fluidic feedback loop underlying the proposed model. Tumor proliferation and diffusive spread act as a volumetric source of interstitial fluid, elevating the local pressure; the resulting pressure gradient drives an outward convective (Darcy) flow that redistributes tumor cells and therapeutic agents, closing the loop back onto subsequent proliferation and transport. Source: Own work and AI.}
\label{fig:concept}
\end{figure}

\ref{fig:concept} summarizes, at a conceptual level, the closed feedback loop that motivates the chemo-fluidic model developed in this work: cellular proliferation and diffusive spreading of the tumor phase act as a volumetric source for the interstitial fluid, elevating the local pressure; the resulting pressure gradient drives an outward convective flow that, in turn, redistributes both tumor cells and dissolved species, feeding back into subsequent proliferation and transport. Capturing this closed loop---rather than proliferation and diffusion in isolation---is the central physical motivation for coupling an advective pressure field to the cellular transport equation.

From a physical standpoint, the pathological elevation of IFP is best understood as an emergent consequence of confined volumetric growth in a saturated poroelastic medium. As the tumor phase expands against the passive resistance of a compressed extracellular matrix and a compromised, structurally abnormal microvasculature, the balance between growth-induced production of interstitial fluid and its clearance through leaky and often collapsed lymphatic channels is disrupted \cite{Jain2014}. As Ambrosi and Mollica \cite{AMBROSI20021297} showed, through a multiplicative decomposition of the deformation gradient into growth and elastic parts, that residual solid stresses accumulate as a direct consequence of geometrically incompatible growth, and that these stresses feed back onto the local proliferation rate through mechanotransduction. This stress-growth feedback loop is one of the central reasons why models that treat $\theta$ in isolation, without any coupling to a mechanical or fluid field, are physically incomplete: they cannot represent the well-documented experimental observation that mechanically confined tumor spheroids exhibit reduced core proliferation relative to their unconfined counterparts. The chemo-fluidic model developed in this work retains the fluid half of this coupling---the part most directly responsible for pressure-driven drug transport---while treating the solid-mechanical half as a modeling simplification whose validity is discussed explicitly in Section \ref{subsec:math_foundation}.

Mathematical oncology has developed models with different levels of complexity. At one extreme, Mixture Theory models \cite{Byrne2003, Preziosi2009} describe tumors as interacting solid and fluid phases and were later extended to poroelastic formulations \cite{Sciume2014} and multi species pharmacological systems \cite{Hu2022}. These models provide detailed biological insight but require many parameters that cannot be reliably estimated from standard clinical imaging, making them difficult to calibrate and apply in practice \cite{Oden2016}.
At the other extreme, reaction diffusion models \cite{Swanson2011, Chen2013, Wong2015} simplify tumor growth by neglecting fluid transport. Although computationally efficient and clinically useful \cite{Swanson2011}, they cannot capture pressure driven transport of cells and therapeutic agents.
A clinically relevant intermediate approach was proposed by Weis et al. \cite{Weis2013}, who incorporated mechanical effects into a reaction diffusion framework using longitudinal imaging data. Their results showed improved prediction of treatment response in breast cancer, highlighting the importance of including mechanical and transport phenomena while maintaining a computationally tractable model. This motivates the intermediate complexity approach adopted in the present work.

Physics-Informed Neural Networks (PINNs), introduced by M. Raissi et al. \cite{RAISSI2019686} as a general framework for solving forward and inverse problems governed by partial differential equations (PDEs), embed the governing residuals directly into the network's loss function through automatic differentiation (AD). This mesh-free approach bypasses the extensive discretization requirements of finite element (FEM) and finite difference (FDM) methods, which is particularly advantageous in geometrically complex domains such as patient-specific tumor geometries reconstructed from medical imaging. 
Beyond geometric flexibility, the PINN paradigm is attractive for inverse problems because the unknown parameters and the unknown continuous fields are estimated jointly within a single differentiable optimization, obviating the need for the repeated forward-solver evaluations that make classical PDE-constrained optimization and Bayesian calibration procedures computationally expensive \cite{Lu2021}. Open-source libraries such as DeepXDE \cite{Lu2021} have substantially lowered the implementation barrier for this class of methods and have been adopted widely across computational science and engineering.

In computational oncology, PINNs have been applied most extensively to reaction-diffusion models of glioblastoma \cite{Subramanian2020, Menze2011}. \cite{Subramanian2020} demonstrated that a PINN trained on sparse longitudinal MRI data could recover patient-specific diffusion and proliferation coefficients, providing a proof of concept for the data-assimilation paradigm in neuro-oncology. The group of Yankeelov \cite{Yankeelov2013} has pursued complementary approaches combining mechanistic PDE models with imaging-derived biomarkers, laying the conceptual groundwork for what the present study formalizes as the DA-PINN framework. A critical obstacle to extending PINNs to richer multiphysical systems is the phenomenon of \emph{gradient pathologies} \cite{Wang2021, Wang2022}. When the loss function comprises residuals from PDEs with disparate differential operators the gradient magnitudes of the respective loss terms can differ by several orders of magnitude, leading the network to only optimizing one of the terms. The resulting ill-conditioned optimization causes the network fail to truly satisfy the physics, producing spurious solutions \cite{Wang2022}. 

The foregoing analysis reveals a clear gap in the existing literature. On the one hand, biophysical fidelity requires incorporating interstitial fluid transport into tumor growth models in order to capture pressure-driven advection and its influence on tumor progression and therapeutic delivery. On the other hand, extending physics-informed neural networks to fully coupled multiphysics formulations introduces severe optimization difficulties arising from the coexistence of governing equations with markedly different mathematical characteristics, leading to the gradient pathologies reported by Wang et al. \cite{Wang2021,Wang2022}. Consequently, there remains a need for a modeling framework that preserves the dominant chemo-fluidic mechanisms while remaining computationally stable and suitable for inverse parameter identification.

The objective of this work is to develop and validate a physics-informed chemo-fluidic modeling framework of intermediate complexity for tumor growth. To this end, we formulate a reduced advection--diffusion--reaction/Darcy model that retains the dominant fluid-mechanical coupling while intentionally omitting the explicit solid-mechanical equilibrium equations under a one-way coupling hypothesis. Based on this formulation, we develop a Physics-Informed Neural Network (PINN) for solving the forward problem and extend it to a Data-Assimilation PINN (DA-PINN) capable of jointly reconstructing the spatiotemporal tumor evolution and estimating hidden transport parameters from sparse observations. The proposed methodology is evaluated through systematic forward and inverse numerical experiments against finite-difference reference solutions.

The remainder of this paper is organized as follows. Section~\ref{sec:math_model} presents the mathematical formulation of the proposed chemo-fluidic model, introduces the simplifying hypotheses, and describes the PINN and DA-PINN formulations together with the training strategy. Section~\ref{sec:results} presents the numerical experiments, including one-dimensional and two-dimensional forward validation against finite-difference solutions and the inverse permeability identification study. Finally, Section~\ref{sec:discussion} discusses the implications of the proposed framework, its current limitations, and potential directions for future research.

\section{Mathematical and Computational Framework}\label{sec:math_model}
This section presents the mathematical and computational framework underlying the proposed model. First, the initial boundary value problem governing tumor growth is formulated. Next, the Physics Informed Neural Network architecture is introduced for solving the forward problem, followed by the Data Assimilation Physics Informed Neural Network framework for the inverse problem and parameter identification.

\subsection{Mathematical Foundation and Hierarchical Derivation}
\label{subsec:math_foundation}

The construction of a computationally tractable yet physically meaningful model for tumor growth requires a rigorous, step-by-step simplification from the most general continuum mechanical description. We trace this hierarchy explicitly to justify each approximation introduced and to clearly delineate the modeling assumptions embedded in our final system.

\subsubsection{Level I: General Mixture Theory (Byrne\cite{Byrne2003} \& Preziosi)\cite{Preziosi2009}}

The most comprehensive continuum descriptions of the TME \cite{Byrne2003,Preziosi2009} treat the tumor as a saturated biphasic mixture in which the solid tumor phase $\theta(\mathbf{x},t) \in [0,1]$ and the interstitial fluid phase $(1-\theta)$ interact through mass and momentum exchange. The theory is built on the classical continuum-mixture postulate that both phases occupy every point of the domain simultaneously, each with its own velocity field, and that the two phases exchange mass (through cell proliferation and death, which convert fluid into new solid volume and vice versa) and momentum (through an interphase drag force analogous to the Darcy resistance of classical poromechanics) \cite{Preziosi2009}. In this framework, the evolution of the tumor volume fraction is governed by

\begin{equation}
\frac{\partial \theta}{\partial t}
+ \nabla \cdot (\theta \mathbf{v}_s)
=
G(\theta,n,\Sigma)
=
\left[
\frac{\gamma}{1+\sigma\Sigma(\theta)}
\frac{(n-\bar{n})_+}{1+\nu n}
-\delta
\right]\theta,
\label{eq:mixture_theta}
\end{equation}

where $\mathbf{v}_s$ is the solid-phase velocity, $n$ is the nutrient concentration, $\Sigma$ is the effective solid stress, $\gamma$ is the stress-modulated proliferation rate, $\sigma$ encodes growth inhibition by mechanical compression, $\nu$ modulates nutrient saturation, $\bar{n}$ is the minimum nutrient threshold for cell viability, and $\delta$ is the apoptosis rate. 

The solid velocity $\mathbf{v}_s$ emerges from the full momentum balance of the mixture, in which the fluid pressure gradient couples to the gradient of effective cellular stresses $\Sigma'(\theta)$ and the tumor viscosities $\lambda_T$ and $\mu_T$:

\begin{equation}
\nabla p
=
-\Sigma'(\theta)\nabla\theta
+ \nabla\!\left(\lambda_T \nabla\cdot\mathbf{v}_s\right)
+ \nabla\cdot\left[
\mu_T
\left(
\nabla\mathbf{v}_s + (\nabla\mathbf{v}_s)^T
\right)
\right].
\label{eq:mixture_momentum}
\end{equation}

Modern extensions of this framework by \cite{Hu2022} add up to 14 coupled ADR equations for individual chemical species $[X_i]$:

\begin{equation}
\frac{\partial [X_i]}{\partial t}
+ \nabla\cdot([X_i]\mathbf{v}_f)
=
\nabla\cdot(D_i\nabla[X_i])
+ f_i([X_1,\dots,X_{14}]),
\label{eq:hu_species}
\end{equation}

where $\mathbf{v}_f$ is the interstitial fluid velocity, $D_i$ is the species-specific diffusivity, and $f_i$ encapsulates nonlinear reaction networks. While mechanistically comprehensive, the resulting system's high dimensionality creates an optimization surface for neural networks that is practically intractable due to competing gradient magnitudes across equations with fundamentally different differential operators \cite{Wang2022}. From an identifiability standpoint, this General Mixture Theory is also poorly suited to sparse clinical data: each additional chemical species introduces further rate constants and diffusivities that are not independently observable from the handful of macroscopic imaging biomarkers (tumor volume fraction, vascular permeability, apparent diffusion coefficient) that are routinely available, so that parameter estimation problems posed directly on Eqs.\ref{eq:mixture_theta}, \ref{eq:mixture_momentum}, \ref{eq:hu_species} are, in practice, severely ill-posed \cite{Oden2016}.

\subsubsection{Level II: Explicit Transport Decoupling (Hu et al.)\cite{Hu2022}}

A significant simplification is achieved by decoupling cellular motion into
(i) intrinsic random motility characterized by an apparent macroscopic diffusivity $D$, and
(ii) convective transport by the interstitial fluid \cite{Hu2022}:

\begin{equation}
\frac{\partial \theta}{\partial t}
=
\underbrace{\nabla\cdot(D\nabla\theta)}_{\text{diffusion}}
-
\underbrace{\nabla\cdot(\theta\mathbf{v}_f)}_{\text{advection}}
+
\underbrace{G(\theta)}_{\text{reaction}}.
\label{eq:level2}
\end{equation}

The advection velocity $\mathbf{v}_f$ is coupled to the interstitial pressure $p$ through Darcy's constitutive law:

\begin{equation}
\mathbf{v}_f = -\kappa\nabla p,
\qquad
\kappa = \frac{K}{\mu_f},
\label{eq:darcy_law}
\end{equation}

where $K$ is the macroscopic tissue permeability and $\mu_f$ is the interstitial fluid viscosity. Darcy's law itself can be understood as the homogenized, pore-scale-averaged limit of the Stokes equations governing viscous fluid flow through the fine, tortuous channels of the interstitial extracellular matrix; the hydraulic conductivity $\kappa$ lumps together the local pore geometry, matrix porosity, and fiber density into a single macroscopic transport coefficient that is, in general, the single most important structural unknown governing convective drug transport, since it directly sets the magnitude of the outward interstitial flow that opposes inward drug perfusion \cite{Jain1988,Pozrikidis2003}. This level retains the dominant fluid-mechanical influence on biology without the full complexity of the Level~I momentum balance.

\subsubsection{Level III: Monophasic Reaction-Diffusion (Chen\cite{Chen2013}, Wong \cite{Wong2015}) , }

The most drastic simplification sets $\mathbf{v}_f = \mathbf{0}$, eliminating the interstitial advection entirely \cite{Chen2013,Wong2015}. The system reduces to a single equation:

\begin{equation}
\frac{\partial \theta}{\partial t}
=
D\nabla^2\theta
+
\rho\theta\!\left(
1 - \frac{\theta}{\theta_{\max}}
\right),
\label{eq:level3}
\end{equation}

which, while tractable, cannot reproduce the pressure-driven transport central to drug delivery resistance and peri-tumoral invasion. It is nonetheless the level of description at which the majority of clinically calibrated glioma models operate \cite{Swanson2011}, precisely because the diffusion coefficient $D$ and proliferation rate $\rho$ of Eq.\ref{eq:level3} are identifiable from as few as two longitudinal MRI acquisitions, illustrating the fundamental trade-off between mechanistic completeness and clinical identifiability that motivates the intermediate-complexity strategy of the present work.

\subsection{Final Proposed IBVP}
\label{subsec:ibvp}

Applying the systematic reductions above to Level~II transport, and replacing the full pressure transient with a quasi-static Darcy source term representative of the interstitial fluid mass balance \cite{Pozrikidis2003,HervasRama2023}, we arrive at the following three-equation coupled system over the domain $\Omega \times [0,T]$:

\begin{align}
    \frac{\partial \theta}{\partial t}
    &= \nabla\cdot(D\nabla\theta) - \nabla\cdot(\theta \mathbf{v}_f)
    + \rho\theta\!\left(
    1-\frac{\theta}{\theta_{\max}}
    \right),
    \label{eq:ibvp_theta}
    \\[6pt]
    \mathbf{v}_f
    &=
    -\kappa \nabla p,
    \label{eq:ibvp_darcy}
    \\[6pt]
    \nabla^2 p
    &=
    -\frac{S}{\kappa}\,\theta(1-\theta),
    \label{eq:ibvp_pressure}
\end{align}

where the right-hand side of Eq.\ref{eq:ibvp_pressure} represents the volumetric fluid production proportional to the net proliferation source $S\,\theta(1-\theta)$, with $S$ the fluid source coefficient. Physically, Eq.\ref{eq:ibvp_pressure} is a quasi-static reduction of the full transient fluid mass balance in which the storage term $\partial p/\partial t$ has been neglected on the grounds that the fluid diffusivity (governed by $\kappa$ and the tissue compressibility) is orders of magnitude larger than the cellular diffusivity $D$; the interstitial pressure field is therefore assumed to re-equilibrate to a quasi-steady Poisson balance on every cellular transport time step, a standard simplification in poromechanical tumor models \cite{Pozrikidis2003}. The primary state variables to be learned by the neural network are therefore $\theta(\mathbf{x},t)$ and $p(\mathbf{x},t)$.

The system is completed by the following boundary and initial conditions over the spatial domain $\Omega$ with boundary $\partial\Omega$ and outward unit normal $\mathbf{n}$:

\begin{itemize}[leftmargin=2em]

\item \textbf{Initial conditions} ($t=0$):
\[
\theta(\mathbf{x},0) = \theta_0(\mathbf{x}),
\qquad
p(\mathbf{x},0)=0,
\]
where $\theta_0(\mathbf{x})$ is a prescribed tumor seed (a Gaussian profile in all experiments) and $p=0$ represents the homeostatic baseline.

\item \textbf{Volume-fraction confinement}:
\[
\bigl(
D\nabla\theta - \theta\mathbf{v}_f
\bigr)\cdot\mathbf{n} = 0
\qquad \text{on } \partial\Omega,
\]
enforcing zero net cellular flux across the domain boundary \cite{Clatz2005}.

\item \textbf{Fluid homeostasis} (Dirichlet):
\[
p = 0
\qquad \text{on } \partial\Omega,
\]
modeling the lymphatic/vascular drainage of the healthy surrounding tissue \cite{Pozrikidis2003}.

\end{itemize}

It should be emphasized that the objective of this work is not to validate the underlying mathematical model itself. Instead, the proposed IBVP adopts a reduced poromechanical formulation that is firmly grounded in the existing literature and closely follows models that have undergone calibration and validation using in vitro, in vivo, and MRI-derived data across multiple tumor types as found in the work of Lorenzo and coworkers \cite{AbbadAndaloussi2026,Lorenzo2024,Lorenzo2024Prostate}. 
The contribution of the present work therefore lies in the development of the PINN framework built upon this established physical model.

\subsection{Physics-Informed Neural Network Formulation} \label{subsec:pinn_formulation}

Having established the governing initial-boundary value problem, the next step is to construct a computational framework capable of solving it efficiently while remaining suitable for inverse parameter identification. To this end, we develop a physics-informed neural network (PINN) that directly embeds the reduced ADR--Darcy equations into the learning process. We first describe the network architecture, then formulate the physics-informed loss, extend the framework to inverse modeling through data assimilation, and finally discuss the modeling assumptions and optimization strategy adopted to ensure stable training.

\subsubsection{Network Architecture}
\begin{figure*}[htbp]
\centering
\includegraphics[width=0.92\textwidth]{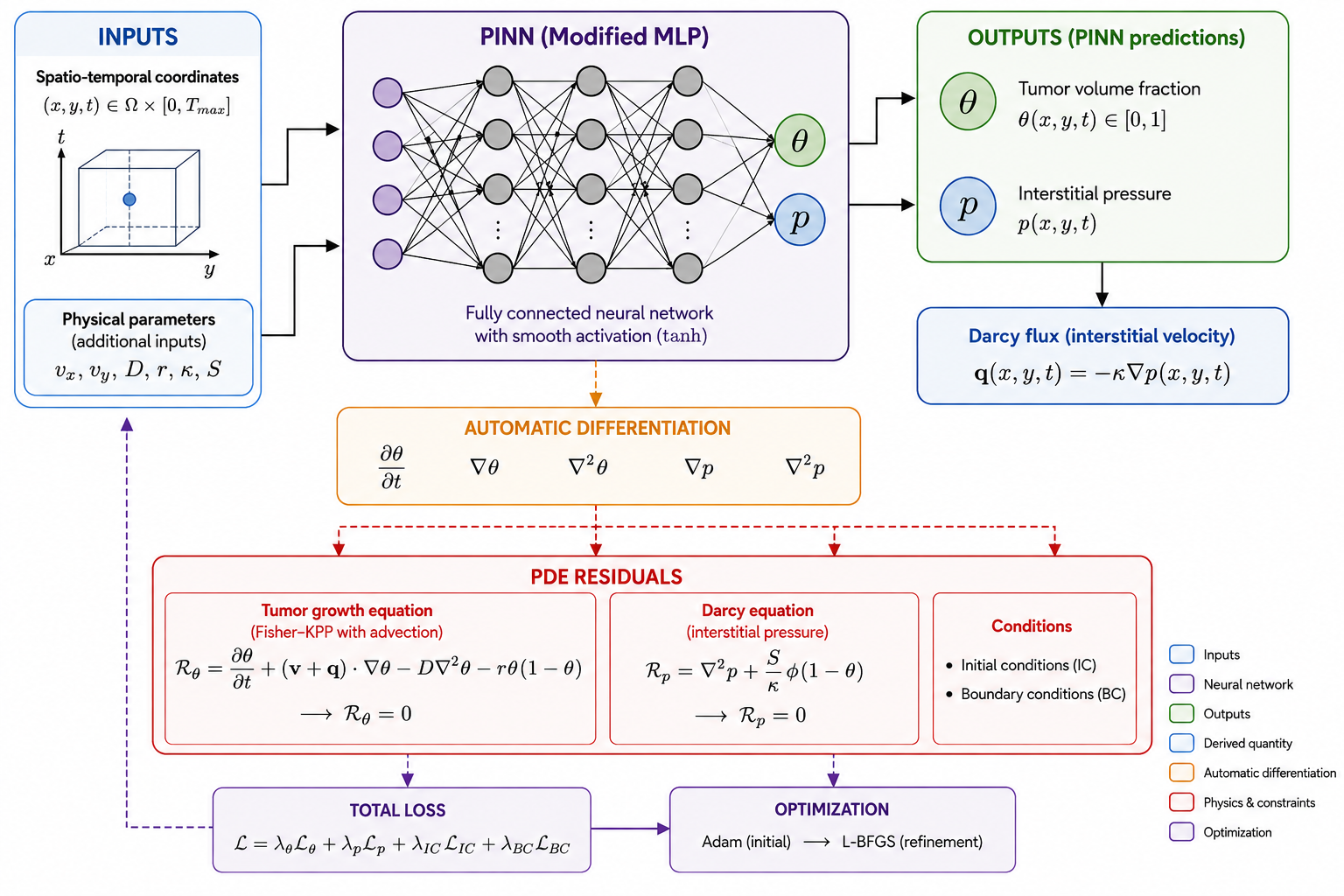}
\caption{End-to-end schematic of the physics-informed learning pipeline. Spatiotemporal coordinates and physical parameters are mapped by a Modified-MLP network to the tumor volume fraction $\theta$ and interstitial pressure $p$; automatic differentiation supplies the derivatives required to evaluate the tumor-growth and Darcy-pressure residuals, which are combined into the total loss and minimized with the hybrid Adam/L-BFGS scheme of \ref{subsec:optimization}. In the inverse (DA-PINN) configuration, the recovered permeability $\kappa$ is fed back as a refined physical parameter, closing the data-assimilation loop. Source: Own work. }
\label{fig:architecture}
\end{figure*}

Let $\mathbf{N} : \mathbb{R}^7 \to \mathbb{R}^2$ denote a fully connected, feed-forward deep neural network parameterized by weights and biases $\theta$. The input vector is $\mathbf{x}_i = (x, y, t, v_x, v_y, \rho, D)^T$, which encodes the spatiotemporal coordinates together with the local biological and hydrodynamic parameters, enabling the network to generalize across the parameter space. The network maps:
\begin{equation}
[\Th, \ph]^T = \mathbf{N}(\mathbf{x}_i;\theta),
\qquad
\Th \in (0,1),\; \ph \geq 0,
\label{eq:net_io}
\end{equation}
where $\widehat{(\cdot)}$ denotes network approximations. Framing the parameters $(v_x, v_y, \rho, D)$ as additional network inputs, rather than as fixed constants baked into the residual, is what allows a single trained network to act as a differentiable surrogate over a neighborhood of the parameter space; this design choice is what subsequently permits the DA-PINN of Section \ref{subsec:inverse_problem} to treat $K$ (and hence $\kappa$) as a quantity to be inferred rather than prescribed, without retraining the network architecture itself \cite{RAISSI2019686}. Figure \ref{fig:architecture} summarizes the complete data flow of the framework, from spatiotemporal and parametric inputs, through the network and its automatic-differentiation-derived operators, to the PDE residuals and the composite loss that drives the Adam/L-BFGS optimization described in \ref{subsec:optimization}.

We adopt the \emph{Modified MLP} architecture of \cite{Wang2022}, in which two auxiliary linear transformations $\mathbf{U}$ and $\mathbf{V}$ act as learned feature gates:
\begin{align}
\mathbf{u} &= \tanh(\mathbf{U}\mathbf{x}_i + \mathbf{b}_U), \notag \\
\mathbf{v} &= \tanh(\mathbf{V}\mathbf{x}_i + \mathbf{b}_V), \notag \\
\mathbf{h}^{(k+1)} &= \tanh\!\bigl(W^{(k)}\mathbf{h}^{(k)} + \mathbf{b}^{(k)}\bigr)
\odot \mathbf{u}
+ \bigl(1 - \tanh(W^{(k)}\mathbf{h}^{(k)} + \mathbf{b}^{(k)})\bigr)
\odot \mathbf{v},
\label{eq:modified_mlp}
\end{align}
for hidden layers $k = 1,\dots,L_h$. Intuitively, the gates $\mathbf{u}$ and $\mathbf{v}$ act as two fixed, learned ``highways'' that are blended into every hidden layer through a convex combination controlled by the layer's own pre-activation; this construction was shown by Wang \cite{Wang2022} to substantially mitigate the vanishing-gradient behavior that otherwise afflicts deep, densely stacked $\tanh$ networks trained against multi-term PDE losses, by ensuring that gradient information from the input layer can propagate to arbitrary depth largely undiminished.

The architecture uses $L_h = 4$ hidden layers with $H = 96$ neurons each, yielding approximately $45{,}000$ trainable parameters.

The output heads enforce physical constraints:
\begin{equation}
\Th = \sigma\!\bigl(f_\theta(\mathbf{h}^{(L_h)})\bigr) \in (0,1),
\qquad
\ph = \mathrm{softplus}\!\bigl(f_p(\mathbf{h}^{(L_h)})\bigr) \geq 0,
\label{eq:output_heads}
\end{equation}
where $\sigma(x)=1/(1+e^{-x})$ is the sigmoid activation and $\mathrm{softplus}(x) = \log(1+e^x)$ enforces non-negativity of pressure. These output parameterizations are a form of \emph{hard constraint} on the admissible solution manifold: rather than penalizing violations of $\theta\in(0,1)$ or $p\geq0$ softly through an additional loss term (which would compete for gradient budget with the PDE residuals and could, in principle, still be violated at convergence), the constraints are satisfied identically by construction for every choice of network weights, which removes two potential sources of gradient-pathology competition from the outset.

All hidden activations use $\tanh$, which is twice continuously differentiable and therefore provides well-defined second-order derivatives required by the PDE residuals.

While the network architecture defines the admissible solution manifold, it does not specify how the parameters are learned. The training process is governed by a physics-informed objective function that penalizes violations of the governing equations together with the prescribed initial and boundary conditions.

\subsubsection{Physics-Informed Loss Function}

The core of the PINN methodology is to penalize violations of the governing PDEs, initial conditions, and boundary conditions at sampled collocation points through a composite loss function:
\begin{equation}
\mathcal{L}(\theta)
= w_{\mathrm{PDE}}\mathcal{L}_{\mathrm{PDE}}
+ w_{\mathrm{IC}}\mathcal{L}_{\mathrm{IC}}
+ w_{\mathrm{BC}}\mathcal{L}_{\mathrm{BC}}.
\label{eq:total_loss_forward}
\end{equation}

with scalar weights $\{w_i\}$ that balance gradient magnitudes between terms.

The physics residual term $\mathcal{L}_{\mathrm{PDE}}$ evaluates the mean squared residuals of Eqs.\ref{eq:res_theta} and \ref{eq:res_p} at $N_f$ collocation points sampled across $\Omega \times [0,T]$:

\begin{equation}
\mathcal{L}_{\mathrm{PDE}}(\theta)
= \frac{1}{N_f}\sum_{i=1}^{N_f}
\bigl[|r_\theta(\mathbf{x}_{f}^i)|^2 + \lambda_p |r_p(\mathbf{x}_{f}^i)|^2\bigr],
\label{eq:loss_pde}
\end{equation}

where the explicit PDE residuals, obtained by substituting \ref{eq:ibvp_darcy} into the governing equations, are:
\begin{align}
r_\theta &:=
\frac{\partial\Th}{\partial t}
+ \bigl(v_x - \kappa\partial_x\ph\bigr)\partial_x\Th
+ \bigl(v_y - \kappa\partial_y\ph\bigr)\partial_y\Th
- D\nabla^2\Th
- \rho\Th(1-\Th),
\label{eq:res_theta}
\\[4pt]
r_p &:=
\partial_{xx}\ph + \partial_{yy}\ph
+ \frac{S}{\kappa}\Th(1-\Th).
\label{eq:res_p}
\end{align}

The initial and boundary condition losses read:
\begin{align}
\mathcal{L}_{\mathrm{IC}} &=
\frac{1}{N_{IC}}\sum_{i=1}^{N_{IC}}
\bigl|\Th(\mathbf{x}_{IC}^i) - \theta_0(\mathbf{x}^i)\bigr|^2,
\\[4pt]
\mathcal{L}_{\mathrm{BC}} &=
\frac{1}{N_{BC}}\sum_{i=1}^{N_{BC}}
\bigl[|\Th(\mathbf{x}_{BC}^i)|^2 + |\ph(\mathbf{x}_{BC}^i)|^2\bigr],
\label{eq:loss_bc}
\end{align}
where the BC loss simultaneously enforces the zero-flux confinement of $\theta$,approximated by a zero Dirichlet condition for simplicity, which is a reasonable proxy given the negligible tumor density at the domain boundary in all experiments, and the fluid homeostasis condition $p=0$ at the domain walls.

\subsubsection{Data-Assimilation Extension (DA-PINN)}

The formulation above addresses the forward problem, where all physical parameters are assumed known. In practical biomedical applications, however, several transport parameters cannot be measured directly and must instead be inferred from sparse clinical observations. This motivates extending the forward PINN into a data-assimilation framework capable of jointly reconstructing the solution fields and estimating unknown model parameters.

For the inverse problem, the loss function is augmented with a data-misfit term:
\begin{equation}
\mathcal{L}^{\mathrm{DA}}(\theta, K)
= \mathcal{L}(\theta)
+ w_{\mathrm{data}}\mathcal{L}_{\mathrm{data}}
+ w_{p}\mathcal{L}_{p,\mathrm{data}},
\label{eq:loss_da}
\end{equation}
where:
\begin{align}
\mathcal{L}_{\mathrm{data}} &=
\frac{1}{N_{\mathrm{obs}}}\sum_{i=1}^{N_{\mathrm{obs}}}
\bigl|\Th(\mathbf{x}_{\mathrm{obs}}^i) - \theta_{\mathrm{obs}}^i\bigr|^2,
\label{eq:loss_data_phi} \\[4pt]
\mathcal{L}_{p,\mathrm{data}} &=
\frac{1}{N_{p}}\sum_{i=1}^{N_p}
\bigl|\ph(\mathbf{x}_{p}^i) - p_{\mathrm{obs}}^i\bigr|^2.
\label{eq:loss_data_p}
\end{align}

Crucially, the macroscopic permeability $K$ is promoted to a scalar \emph{trainable parameter} of the optimization problem:

\begin{equation}
K = \exp(K_{\log}),
\qquad K_{\log} \in \mathbb{R},
\label{eq:K_log}
\end{equation}
where the log-parameterization enforces $K > 0$ and improves optimization stability by transforming the strictly positive constraint into an unconstrained problem \cite{RAISSI2019686}. This construction means that the gradient of the loss with respect to $K_{\log}$ flows through both the PDE residuals (via $\kappa = K/\mu_f$ in $r_\theta$ and $r_p$) and the data-misfit terms, providing a rich gradient signal for parameter identification. The log-parameterization additionally has the effect of equalizing the relative step size of the optimizer across orders of magnitude of $K$: because $\mathrm{d}K/K = \mathrm{d}K_{\log}$, a fixed learning rate on $K_{\log}$ produces geometric, rather than arithmetic, updates on $K$ itself, which is well matched to the $4\times$ initial misspecification of $K_{\mathrm{init}}$ relative to $K_{\mathrm{true}}$ considered in Section \ref{subsec:inverse_problem}, and is a standard device in Bayesian and variational estimation of strictly positive physical constants.

\paragraph{Dedicated Darcy Constitutive Residual}

A key insight of this work is that $K$ only enters the forward-problem residuals implicitly through the Darcy velocity $\bq = -\kappa\nabla p$. Under a naive loss weighting, the network can satisfy the tumor PDE by absorbing the Darcy correction into the background velocity field $\mathbf{v}_0$, rendering $K$ \emph{unidentifiable} from $\theta$-observations alone. This is a textbook instance of a \emph{structural} (as opposed to \emph{practical}) identifiability failure: even in the idealized limit of noise-free, dense observations of $\theta$ alone, the composite advective velocity $v_x - \kappa\partial_x p$ appearing in Eq.\ref{eq:res_theta} constrains only the product of $\kappa$ and $\nabla p$, so that infinitely many pairs $(\kappa, p)$ reproduce an identical $\theta$ field, and the optimizer is free to select whichever pair happens to minimize the loss fastest---which, absent additional information, need not be the physically correct one. To break this degeneracy, we introduce an additional \emph{Darcy constitutive residual} that explicitly constrains the flux divergence:
\begin{equation}
r_{\mathrm{Darcy}} :=
\nabla\cdot\bq + \frac{S}{\kappa}\Th(1-\Th)
= -\kappa\nabla^2\ph + \frac{S}{\kappa}\Th(1-\Th),
\label{eq:res_darcy_constitutive}
\end{equation}
with corresponding loss term $\mathcal{L}_{\mathrm{Darcy}} = \|r_{\mathrm{Darcy}}\|_2^2 / N_f$, weighted at $w_{\mathrm{Darcy}} = 15$ in the inverse problem. This term forces $K$ to matter at \emph{every} Darcy collocation point, ensuring that the permeability gradient is non-negligible throughout training, and it is complemented by the sparse pressure observations $\mathcal{L}_{p,\mathrm{data}}$, which anchor the absolute scale of $p$ (and hence of $\kappa$) that the flux-only residual of Eq.\ref{eq:res_theta} cannot fix on its own.

\subsection{Hypotheses and Rationale for Mechanical Decoupling}
\label{subsec:decoupling}

Having established the computational framework, it remains to justify one of the principal modeling decisions adopted throughout this work: the omission of the explicit solid-mechanical equilibrium equations. The following subsection discusses the physical and numerical arguments supporting this reduced formulation.

The reduced ADR--Darcy formulation proposed in Section~\ref{subsec:ibvp} intentionally omits the explicit equilibrium equations governing solid tissue deformation. Since this simplification fundamentally determines both the mathematical model and the subsequent PINN formulation, its physical justification deserves explicit discussion. The rationale rests on two complementary arguments: a biophysical hypothesis regarding the dominant transport mechanisms and a numerical consideration related to the optimization of multi-physics PINNs.

\subsubsection{Biophysical one-way coupling.}

For infiltrating tumors with rapid cellular expansion, macroscopic tissue deformation is a secondary effect compared to the advective and diffusive timescales of cellular transport \cite{Clatz2005}. We hypothesize that the mechanical feedback of solid stress into the growth function can be neglected for the identification of transport parameters, consistent with the one-way coupling assumption of \cite{Chen2013}. This hypothesis can be framed formally as a time-scale separation argument: denoting by $\tau_{\mathrm{mech}}$ the characteristic relaxation time of the viscoelastic tumor tissue and by $\tau_{\mathrm{transport}} = L^2/D$ the characteristic diffusive time over the domain length scale $L$, the one-way coupling approximation is expected to hold whenever $\tau_{\mathrm{mech}} \ll \tau_{\mathrm{transport}}$, so that the solid skeleton effectively equilibrates instantaneously on the transport time scale and its explicit dynamics can be absorbed into a quasi-static background velocity field $\mathbf{v}_0$ rather than solved for explicitly \cite{Clatz2005}. We emphasize that this is a modeling hypothesis whose regime of validity is tumor-type and stage dependent: it is best justified for diffusively infiltrating lesions such as glioblastoma, and less so for tumors with pronounced desmoplastic, stiff stromal reactions in which stress-growth feedback of the kind identified by \cite{AMBROSI20021297} is a first-order effect rather than a secondary correction.
Although this assumption is physically motivated, it also offers an important computational advantage when solving the resulting inverse problem with PINNs.

\subsubsection{Numerical stability and gradient pathologies.}
Incorporating the full Biot system of solid equilibrium:

\begin{equation}
-\!\left(K_u - \tfrac{2G}{3}\right)\!\nabla(\nabla\cdot\mathbf{u})
- G\nabla^2\mathbf{u}
+ \alpha\nabla p
+ K\nabla\theta
=
\mathbf{0},
\label{eq:biot}
\end{equation}

where $K_u$ and $G$ are the undrained bulk and shear moduli and $\alpha$ is the Biot--Willis coefficient, introduces high-order spatial operators and an additional state variable $\mathbf{u}$. For a PINN optimizer, the second-order elliptic nature of Eq. \ref{eq:biot} produces gradient magnitudes that systematically overwhelm those of the parabolic transport residual, leading to the gradient pathologies. This can be understood quantitatively through the NTK argument of Wang et al. \cite{Wang2022}: the eigenvalues of the kernel associated with the Biot operator in Eq. \ref{eq:biot} scale with a higher power of spatial frequency than those associated with the parabolic residual of Eq. \ref{eq:ibvp_theta}, so that, under a fixed, unweighted gradient-descent schedule, the elastic residual is minimized at a systematically faster rate than the transport residual, starving the latter of effective training signal long before the network has had the opportunity to resolve the tumor field itself. By removing this subsystem, we transform the loss landscape into one that is numerically stable and amenable to the hybrid Adam/L-BFGS training strategy described in \ref{subsec:pinn_formulation}, at the cost of forfeiting direct access to the solid-stress field and to any stress-mediated growth inhibition, a limitation we return to in Section \ref{sec:discussion}.

\subsubsection{Collocation Strategy and Optimization}
\label{subsec:optimization}

With the governing equations, loss function and modeling assumptions established, the remaining task is to solve the resulting optimization problem efficiently. The training procedure combines stochastic collocation with a hybrid first- and second-order optimization strategy designed to balance global exploration and local convergence.

Collocation points are sampled uniformly at random within $\Omega \times [0,T]$ at each training iteration (online sampling), which provides stochastic regularization and avoids overfitting to specific spatial regions. The optimization employs a \emph{hybrid two-phase strategy}:
\begin{enumerate}[leftmargin=2em]
\item \textbf{Adam phase} \cite{Kingma2015}: $N_{\mathrm{Adam}}$ stochastic gradient steps with cosine annealing learning rate $\eta_0 = 2\times10^{-3} \to \eta_{\min} = 10^{-4}$. For the DA-PINN, the network and $K_{\log}$ are assigned \emph{separate} Adam instances, with the $K_{\log}$ optimizer using $\eta_K = 10^{-2}$ (5$\times$ higher). A warm-up period of $N_{\mathrm{wu}} = 1{,}000$ iterations freezes $K_{\log}$ to allow the network to first learn a physically consistent solution before permeability identification begins.
\item \textbf{L-BFGS phase}: $N_{\mathrm{LBFGS}}$ quasi-Newton steps using the strong Wolfe line search \cite{Liu1989}, providing high-accuracy local convergence after Adam has navigated the global landscape.
\end{enumerate}
To mitigate gradient pathologies, the loss weight vector $\mathbf{w} = (w_\theta, w_p, w_{\mathrm{Darcy}}, w_{\mathrm{IC}}, w_{\mathrm{BC}}, w_{\mathrm{data}}, w_{p,\mathrm{data}})$ is set as reported in Table \ref{tab:loss_weights} based on a manual calibration procedure: we adjust weights until the time-averaged gradient norms of each loss component are within one order of magnitude of each other, following the recommendation of \cite{Wang2021}. The warm-up freeze of $K_{\log}$ during the first $N_{\mathrm{wu}}$ Adam iterations plays a complementary role to the loss-weight calibration: it sequences the optimization so that the network first descends toward a state-consistent solution manifold under a fixed (if initially incorrect) permeability, and only then begins to perturb $K_{\log}$, which empirically prevents the joint optimization from being trapped in the shallow, non-physical local minima that a simultaneous cold start of both the network weights and $K_{\log}$ is prone to produce. \ref{fig:training_convergence} shows a representative training trace for the forward problem, in which the total loss and its four constituent components (initial condition, supervised data where applicable, PDE residual, and boundary condition) decrease over several orders of magnitude, with the visible plateaus corresponding to learning-rate decay steps in the cosine annealing schedule.

\begin{table}[htbp]
\centering
\caption{Loss function weights for the forward PINN (columns 2--3) and DA-PINN inverse problem (columns 4--5).}
\label{tab:loss_weights}
\begin{tabular}{lcccc}
\toprule
Term & $w$ (Forward) & Rationale & $w$ (Inverse) & Rationale \\
\midrule
$w_\theta$ (PDE $\theta$) & 1.0 & baseline & 1.0 & unchanged \\
$w_p$ (PDE $p$)           & 0.5 & secondary & 8.0 & $K$ identifiability \\
$w_{\mathrm{Darcy}}$      & ---   & not used & 15.0 & $K$ load-bearing \\
$w_{\mathrm{IC}}$         & 10.0 & strong IC & 10.0 & unchanged \\
$w_{\mathrm{BC}}$         & 5.0  & moderate & 5.0 & $+p$ BC \\
$w_{\mathrm{data}}$ ($\theta$) & --- & not used & 20.0 & data fit \\
$w_{p,\mathrm{data}}$ ($p$)    & --- & not used & 10.0 & $K$ identifiability \\
\bottomrule
\end{tabular}
\end{table}


\section{Numerical Experiments and Results} \label{sec:results}

This section presents the numerical validation of the proposed framework through three experiments. We first evaluate the PINN on a one dimensional forward problem to verify its ability to reproduce the governing physics under a simplified setting. Next, the methodology is extended to the full two dimensional forward problem, demonstrating its performance for the coupled tumor growth model. Finally, the Data Assimilation PINN is applied to the inverse problem, where the permeability is identified from synthetic observations.
All simulations and neural network models were implemented in a unified Python framework. Reference solutions for the forward problems were generated using a finite difference solver developed specifically for this work. To accelerate both the numerical simulations and the PINN training, the implementation was parallelized and executed on GPU hardware using automatic differentiation and optimized tensor operations.
A summary of the physical parameters used throughout the numerical experiments is provided in Table~\ref{tab:params}.

\begin{table}[htbp]
\centering
\caption{
Physical and numerical parameters used in all experiments.
The parameter marked with $(\star)$ is the target of inverse recovery in
\ref{subsec:inverse_problem}.
}
\label{tab:params}

\begin{tabular}{llll}
\toprule
Symbol & Description & Value & Units \\
\midrule

$L$
& Domain side length
& 10
& \si{\milli\metre}
\\

$T$
& Simulation horizon
& 6
& \si{\day}
\\

$K^\star$
& Macroscopic permeability
& 0.02
& \si{\milli\metre^2\per\pascal\per\day}
\\

$\mu_f$
& Fluid viscosity
& 1.0
& \si{\pascal\cdot\day}
\\

$\kappa = K/\mu_f$
& Hydraulic conductivity
& 0.02
& \si{\milli\metre^2\per\day}
\\

$S$
& Fluid source coefficient
& 0.05
& \si{\per\pascal}
\\

$D$
& Cellular diffusivity
& 0.10
& \si{\milli\metre^2\per\day}
\\

$R,\rho$
& Proliferation rate
& 0.30
& \si{\per\day}
\\

$v_{x,0}$
& Background fluid velocity in $x$
& 0.30
& \si{\milli\metre\per\day}
\\

$v_{y,0}$
& Background fluid velocity in $y$
& 0.20
& \si{\milli\metre\per\day}
\\

$\theta_0$
& Initial tumor center (Gaussian $\sigma=0.6$)
& 1
& ---
\\

\bottomrule
\end{tabular}

\end{table}

\subsection{1D Forward Problem: Space-Time Verification}
\label{subsec:1d_forward}

As an initial validation of the proposed framework, we consider the one-dimensional reduction of Eqs.~\ref{eq:ibvp_theta}, \ref{eq:ibvp_darcy}, \ref{eq:ibvp_pressure}. The objective of this experiment is to verify that the forward PINN accurately reproduces the solution of the governing advection--diffusion--reaction equation before addressing the full two-dimensional coupled problem. A finite-difference (FD) solver was implemented as the reference numerical solution, against which the PINN predictions were compared.

The computational domain spans $x\in[0,10]$ mm over a simulation horizon of $T=8$ days. The FD solution is computed on a uniform grid comprising $N_x=100$ spatial nodes and $N_t=401$ temporal steps. The PINN receives the spatiotemporal coordinates together with the governing physical parameters $(v,\rho,D)$ as inputs and predicts the tumor volume fraction $\theta(x,t)$. Table~\ref{tab:1d_validation} summarizes the numerical configuration employed throughout this validation study.

\begin{table}[htbp]
\centering
\caption{Configuration of the one-dimensional forward validation experiment.}
\label{tab:1d_validation}
\begin{tabular}{ll}
\toprule
Quantity & Value\\
\midrule
Spatial domain & $x\in[0,10]$ mm\\
Simulation time & $t\in[0,8]$ days\\
Spatial discretization & $N_x=100$\\
Temporal discretization & $N_t=401$\\
Reference solver & Explicit finite differences\\
PINN inputs & $(x,t,v,\rho,D)$\\
PINN output & Tumor volume fraction $\theta(x,t)$\\
Comparison metrics & MAE, RMSE, Relative $L_2$, Maximum error\\
\bottomrule
\end{tabular}
\end{table}

To assess the robustness of the model under different transport regimes, three representative parameter combinations were considered. The first corresponds to a moderate balance between advection and diffusion, the second increases the effective diffusion coefficient, and the third introduces a stronger background velocity together with reduced diffusivity, producing a sharper propagating front. These cases are summarized in Table~\ref{tab:1d_cases}.

\begin{table}[htbp]
\centering
\caption{Transport parameters used in the one-dimensional validation cases.}
\label{tab:1d_cases}
\begin{tabular}{cccc}
\toprule
Case & $v$ & $\rho$ & $D$\\
\midrule
Case 1 & 1.0 & 0.30 & 0.12\\
Case 2 & 1.0 & 0.30 & 0.15\\
Case 3 & 1.2 & 0.20 & 0.05\\
\bottomrule
\end{tabular}
\end{table}

\begin{figure}[htbp]
    \centering
    \includegraphics[width=\linewidth]{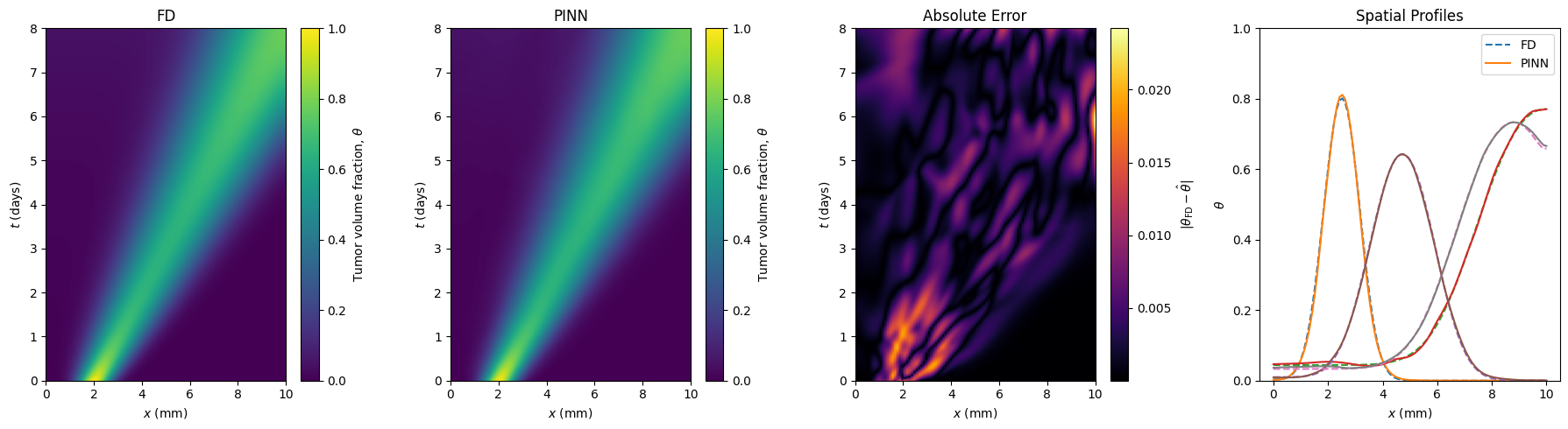}
    \caption{One-dimensional forward validation for Case 1. From left to right: finite-difference (FD) reference solution, PINN prediction, pointwise absolute error, and representative spatial profiles at several time instants. In the first three panels, the horizontal axis corresponds to the spatial coordinate $x$ (mm) and the vertical axis to time $t$ (days). In the rightmost panel, the horizontal axis is the spatial coordinate $x$ (mm) and the vertical axis the tumor volume fraction $\theta$. Solid lines denote the PINN prediction and dashed lines the FD solution.}
    \label{fig:1d_case1}
\end{figure}

\begin{figure}[htbp]
    \centering
    \includegraphics[width=\linewidth]{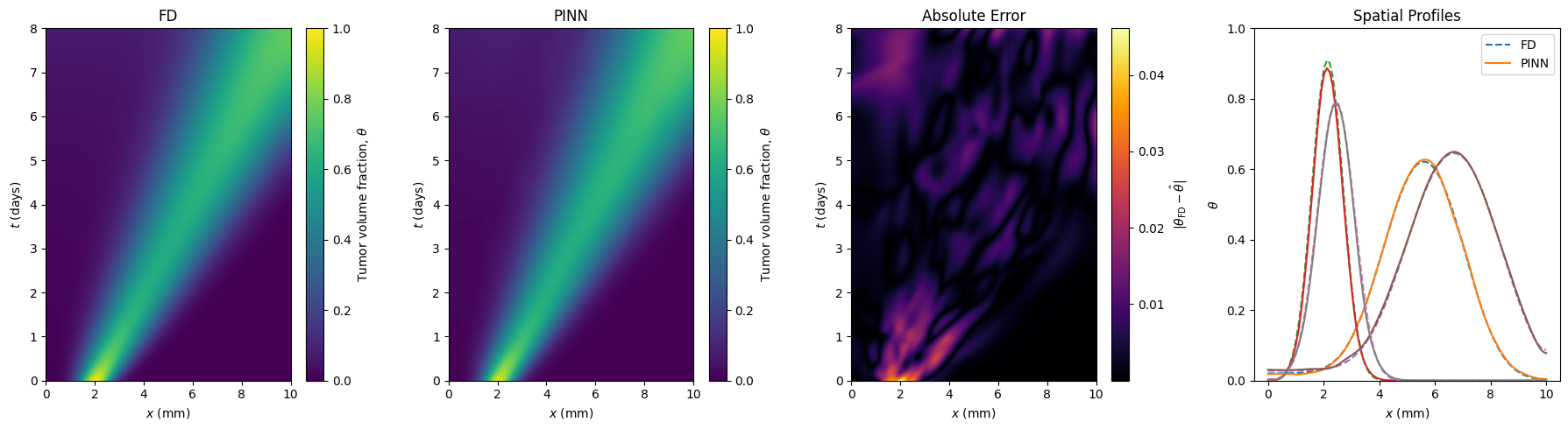}
    \caption{One-dimensional forward validation for Case 2. Panel layout is identical to Fig.~\ref{fig:1d_case1}. Increasing the diffusion coefficient produces a broader propagating front while maintaining excellent agreement between the PINN and the finite-difference reference.}
    \label{fig:placeholder}
\end{figure}

\begin{figure}[htbp]
    \centering
    \includegraphics[width=\linewidth]{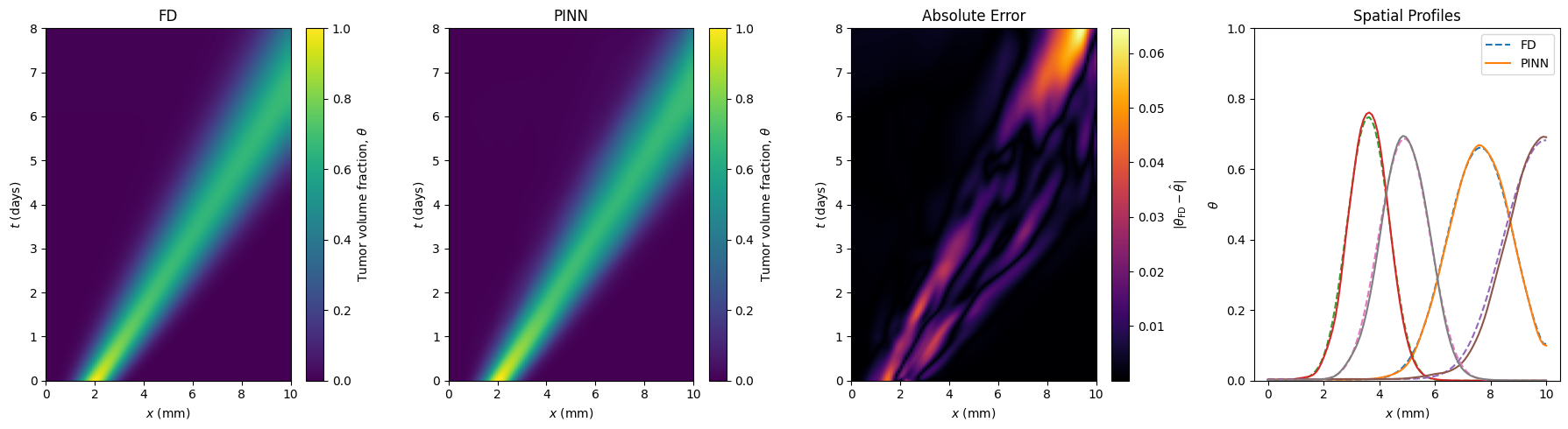}
    \caption{One-dimensional forward validation for Case 3. Panel layout is identical to Fig.~\ref{fig:1d_case1}. The increased advection velocity and reduced diffusion produce a sharper travelling front, resulting in a slightly larger localized error near the advancing interface while preserving excellent agreement over the remainder of the computational domain.}
    \label{fig:1d_case3}
\end{figure}

Figures~\ref{fig:1d_case1}--\ref{fig:1d_case3} compare the PINN solution with the FD reference for the three test cases. In each figure, the first column shows the FD solution, the second the PINN prediction, the third the pointwise absolute error, and the fourth representative spatial profiles at several time instants. For the first three panels, the horizontal axis corresponds to the spatial coordinate $x$ (mm), while the vertical axis represents time $t$ (days). The rightmost panel depicts the tumor volume fraction as a function of spatial position, where solid lines denote the PINN prediction and dashed lines the FD reference.

Across all cases, the PINN accurately reproduces the propagating Fisher--KPP travelling wave generated by the FD solver. The largest discrepancies are consistently localized around the advancing tumor front, where the solution exhibits the steepest spatial gradients, whereas the agreement is nearly indistinguishable elsewhere in the computational domain. The spatial profiles further demonstrate that the network correctly captures both the propagation speed and the morphology of the tumor front throughout the simulation.

\newpage

To quantify the agreement, Table~\ref{tab:1d_errors} reports the mean absolute error (MAE), root-mean-square error (RMSE), relative $L_2$ error, and maximum pointwise error for each case. All three experiments exhibit relative $L_2$ errors below $5\%$, confirming that the proposed PINN accurately approximates the reference finite-difference solution over a range of transport conditions.

\begin{table}[htbp]
\centering
\caption{Quantitative comparison between the PINN and FD solutions for validation cases.}
\label{tab:1d_errors}
\begin{tabular}{ccccc}
\toprule
Case & MAE & RMSE & Relative $L_2$ & Maximum error\\
\midrule
1 & $5.834\times10^{-3}$ & $8.120\times10^{-3}$ & $2.437\times10^{-2}$ & $3.606\times10^{-2}$\\
2 & $6.485\times10^{-3}$ & $9.222\times10^{-3}$ & $2.778\times10^{-2}$ & $6.331\times10^{-2}$\\
3 & $6.182\times10^{-3}$ & $1.151\times10^{-2}$ & $4.232\times10^{-2}$ & $6.314\times10^{-2}$\\
\bottomrule
\end{tabular}
\end{table}

\subsection{2D Forward Problem: PINN vs. Finite Differences}
\label{subsec:2d_forward}

Having verified the one-dimensional formulation, we next consider the full two-dimensional ADR--Darcy system to evaluate whether the proposed PINN can accurately reproduce the coupled tumor-growth dynamics in a more realistic setting. Throughout this section, the finite-difference (FD) solver serves as the numerical reference solution, while the PINN is assessed in terms of its ability to reconstruct the spatiotemporal tumor field, recover clinically relevant derived quantities, and generalize across different transport regimes.

Unless otherwise stated, the network is trained using the optimization strategy described in Section~\ref{subsec:optimization}, consisting of $N_{\mathrm{Adam}}=5\,000$ Adam iterations followed by $N_{\mathrm{LBFGS}}=300$ L-BFGS iterations. At each optimization step, $N_f=512$ interior collocation points are sampled uniformly over $\Omega\times[0,T]$, together with $N_{IC}=512$ initial-condition points and $N_{BC}=256$ boundary-condition points.

\subsubsection{Field Reconstruction}

\begin{figure*}[htbp]
\centering
\includegraphics[width=0.98\textwidth]{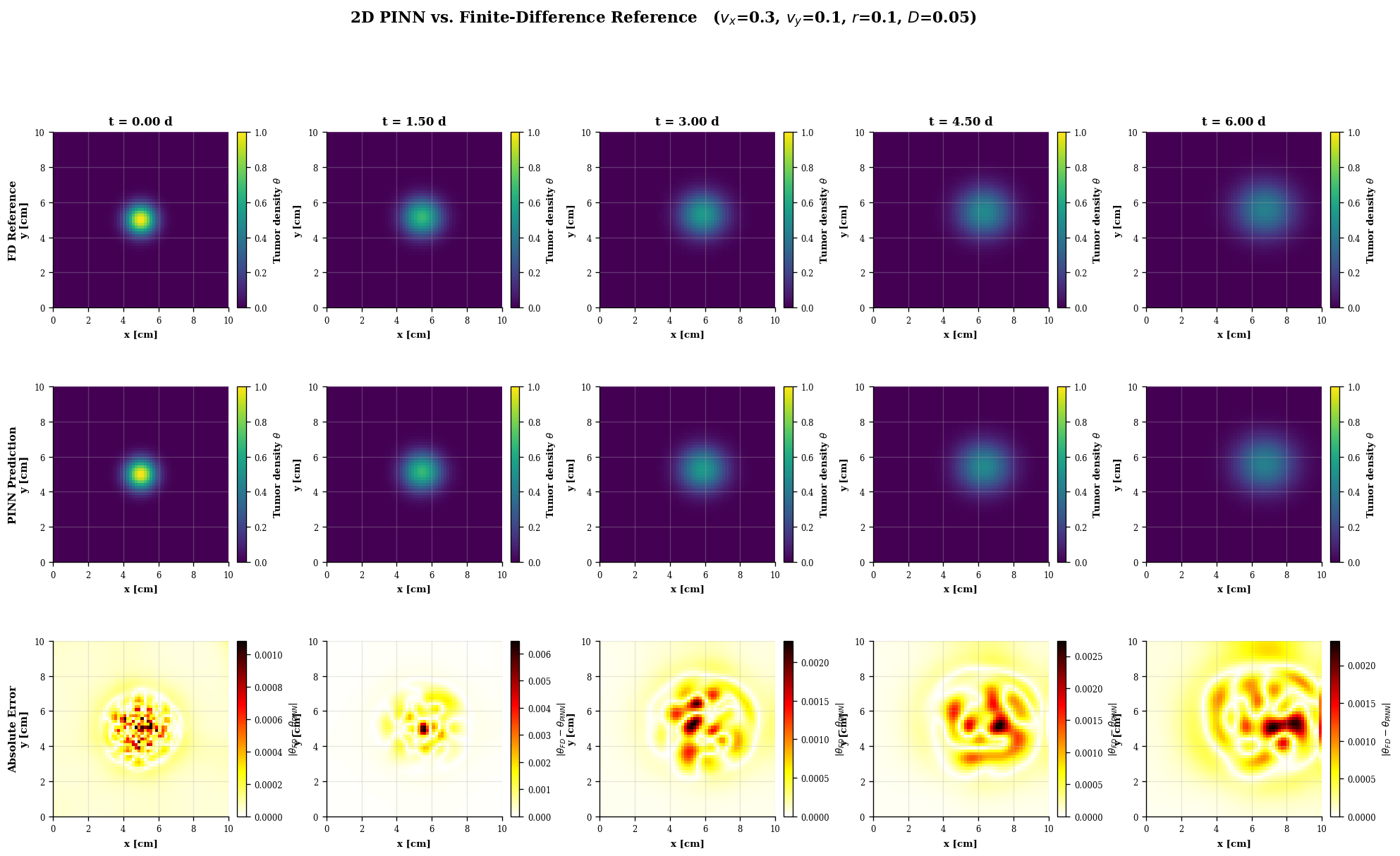}
\caption{Two-dimensional forward-problem verification for the baseline parameter set ($v_x=0.3$, $v_y=0.1$, $\rho=0.1$, $D=0.05$). Top row: finite-difference reference tumor volume fraction $\theta_{\mathrm{FD}}(x,y,t)$ at five time instants. Middle row: PINN prediction $\Th(x,y,t)$. Bottom row: pointwise absolute error $|\theta_{\mathrm{FD}} - \Th|$, with a per-panel color scale. The tumor seed spreads outward under combined diffusion, logistic proliferation, and advection by the background fluid velocity, producing the visible asymmetric drift toward the upper-right of the domain.}
\label{fig:phi_evolution}
\end{figure*}

We first evaluate the ability of the network to recover the complete spatiotemporal tumor field. Figure~\ref{fig:phi_evolution} compares the PINN prediction with the finite-difference reference for the baseline parameter set reported in Table~\ref{tab:params} ($v_x=0.3$, $v_y=0.1$, $\rho=0.1$, $D=0.05$). Snapshots are shown at five equally spaced time instants,
$t=\{0,1.5,3.0,4.5,6.0\}$ days.

The first row presents the FD reference solution, the second row the corresponding PINN prediction, and the third row the pointwise absolute error. In every panel, the horizontal and vertical axes represent the spatial coordinates $x$ and $y$ (mm), respectively. The colour scale denotes the tumor volume fraction $\theta$ for the first two rows and the absolute error $|\theta_{\mathrm{FD}}-\widehat{\theta}|$ for the third.

The PINN accurately reproduces the Gaussian initial condition together with the subsequent outward expansion of the tumor driven by diffusion, logistic proliferation, and background advection. The asymmetric displacement of the tumor towards the upper-right corner of the domain, induced by the imposed interstitial velocity, is correctly captured throughout the simulation. The error remains below $2.5\times10^{-3}$ over the entire time horizon and is primarily concentrated along the advancing tumor interface, where the solution exhibits its largest spatial gradients.

\begin{figure*}[htbp]
\centering
\includegraphics[width=\textwidth]{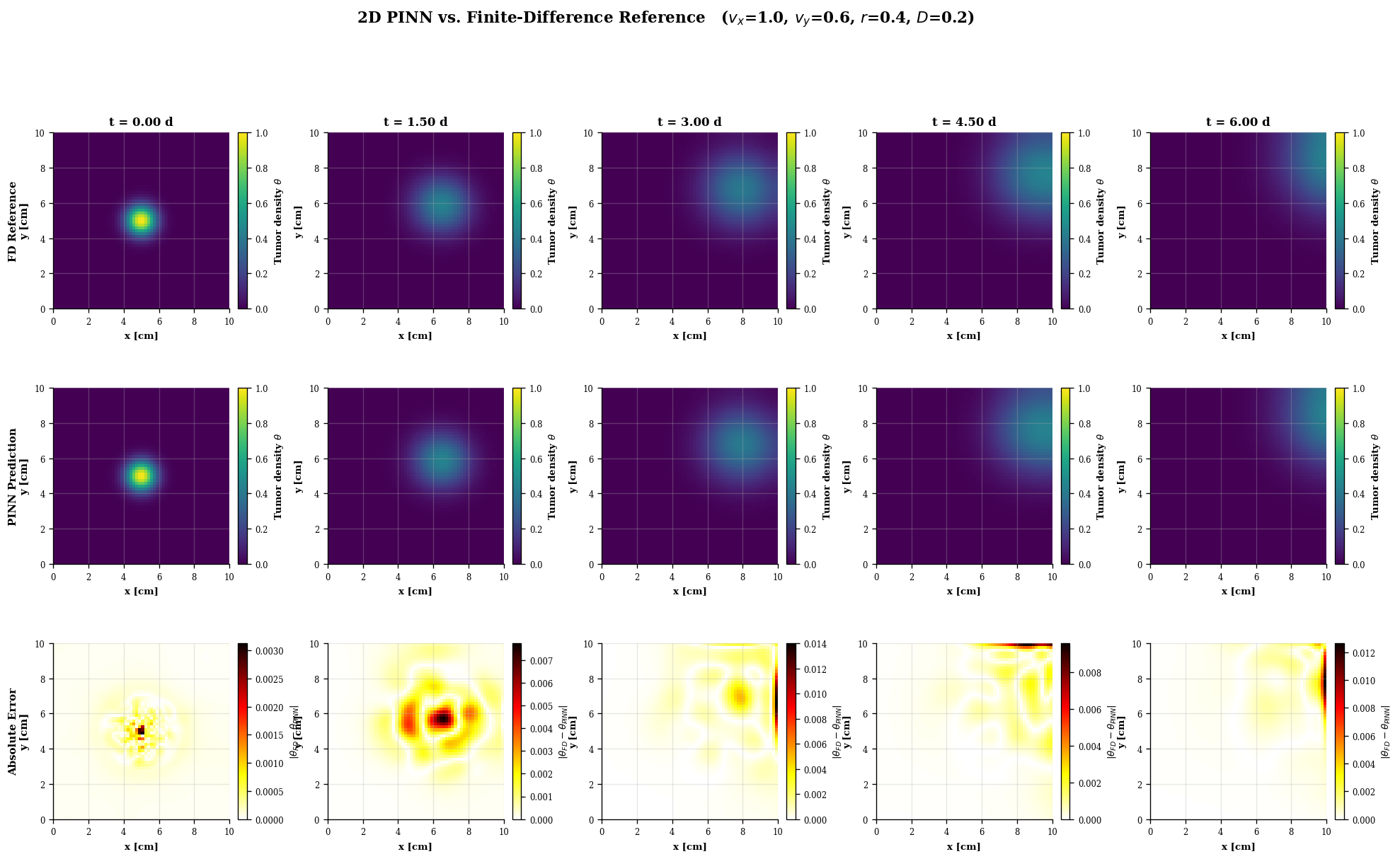}
\caption{Two-dimensional forward-problem verification for a faster-growing, more strongly advected parameter set ($v_x=1.0$, $v_y=0.6$, $\rho=0.4$, $D=0.2$). Panel layout as in \ref{fig:phi_evolution}. The larger proliferation and advection coefficients produce a tumor front that reaches the domain boundary within the simulated horizon; the largest pointwise errors are correspondingly concentrated near the boundary at later times, where the fixed-domain confinement condition interacts with the fast-moving front.}
\label{fig:phi_evolution_faster}
\end{figure*}

To evaluate the robustness of the proposed framework, the experiment is repeated for a substantially more challenging transport regime characterized by stronger advection and faster proliferation ($v_x=1.0$, $v_y=0.6$, $\rho=0.4$, $D=0.2$). The resulting comparison is presented in Fig.~\ref{fig:phi_evolution_faster}.

Despite the markedly different physical parameters, the PINN continues to closely reproduce the FD solution throughout most of the computational domain. The largest discrepancies appear only during the final stages of the simulation, when the rapidly advancing tumor front reaches the fixed domain boundaries. In this regime, the interaction between the imposed Dirichlet pressure condition and the confined computational domain produces localized errors near the upper and right boundaries. Importantly, these discrepancies remain confined to the boundary layer and do not affect the reconstruction of the interior tumor field, demonstrating that the proposed architecture generalizes well across substantially different transport regimes without requiring architectural modifications.

\subsubsection{Derived Diagnostic Quantities}

\begin{figure}[htbp]
\centering
\includegraphics[width=\linewidth]{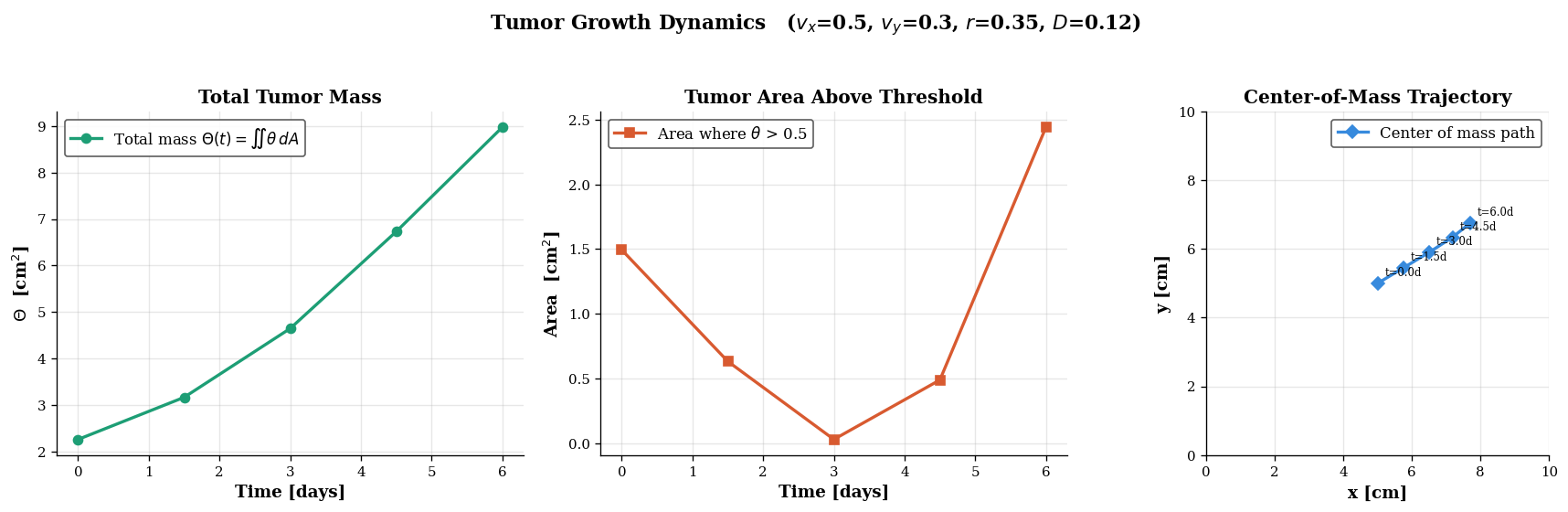}
\caption{Macroscopic quantities derived from the PINN-predicted tumor field. Left: evolution of the total tumor mass,
$\Phi(t)=\int_\Omega \theta\,\mathrm{d}A$. Center: tumor area above the detection threshold ($\theta>0.5$), used as a surrogate for the visible lesion area. Right: trajectory of the tumor center of mass, illustrating the net displacement induced by background interstitial advection.}
\label{fig:tumor_dynamics}
\end{figure}

Beyond reproducing the tumor field itself, the proposed PINN enables the computation of clinically relevant quantities derived from the predicted solution. Figure~\ref{fig:tumor_dynamics} summarizes three representative macroscopic diagnostics computed from the tumor volume fraction. The left panel reports the total tumor mass,
\[
\Phi(t)=\int_{\Omega}\theta\,\mathrm{d}A,
\]
which increases monotonically throughout the simulation as a consequence of sustained logistic proliferation. The central panel shows the tumor area above a detection threshold ($\theta>0.5$), used here as a simple surrogate for the visible lesion that would be segmented from a thresholded diffusion-weighted MRI image. Unlike the total mass, this quantity exhibits an initial decrease as the compact tumor seed spreads and locally dilutes below the threshold, followed by renewed growth as proliferation dominates. Finally, the right panel illustrates the trajectory of the tumor center of mass, which drifts in the direction of the imposed background interstitial velocity. This displacement provides a clear macroscopic signature of advective transport, a behavior that cannot be reproduced by purely reaction--diffusion models such as Eq.~\ref{eq:level3}.

\begin{figure}[htbp]
\centering
\includegraphics[width=\linewidth]{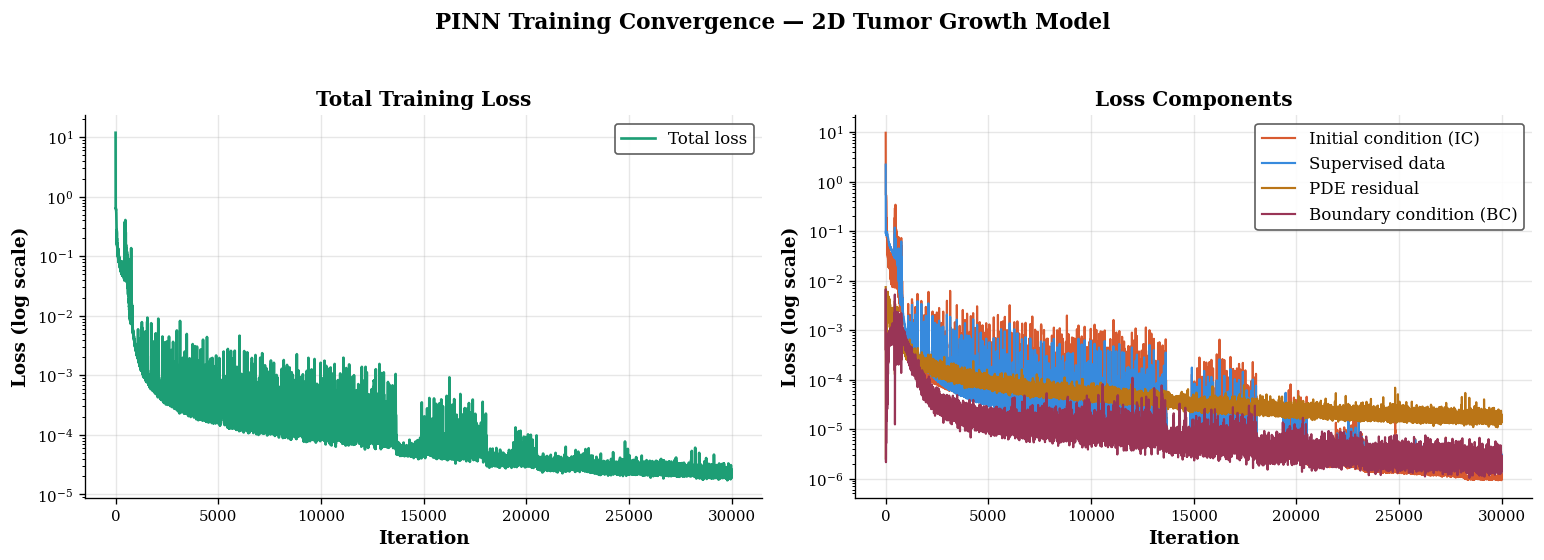}
\caption{Representative training history of the forward PINN. Left: evolution of the total loss during the hybrid Adam/L-BFGS optimization. Right: evolution of the individual loss components, showing balanced convergence of the initial-condition, boundary-condition and PDE residual terms after applying the calibrated loss weights of Table~\ref{tab:loss_weights}.}
\label{fig:training_convergence}
\end{figure}

The convergence behaviour of the proposed optimization strategy is illustrated in Fig.~\ref{fig:training_convergence}. The left panel presents the evolution of the total loss throughout training, showing a reduction of approximately six orders of magnitude before convergence. The short plateaus correspond to the scheduled learning-rate reductions during the Adam phase, after which the L-BFGS optimizer refines the solution. The right panel reports the individual loss components, demonstrating that the initial-condition, boundary-condition, and PDE residual losses decrease simultaneously without any single term dominating the optimization. This balanced convergence confirms the effectiveness of the loss-weight calibration described in Section~\ref{subsec:optimization} for mitigating the gradient pathologies commonly encountered in multi-physics PINNs.

\subsubsection{Quantitative Validation}

Although the previous figures demonstrate excellent qualitative agreement, a quantitative comparison with the finite-difference solution is required to assess the predictive accuracy of the proposed surrogate. Table~\ref{tab:fd_comparison} summarizes several standard error metrics evaluated at the midpoint of the simulation ($t=T/2=3$ days).

\begin{equation}
  e_{L^2}^{\mathrm{rel}}
  = \frac{\|\Th - \theta_{\mathrm{FD}}\|_{L^2(\Omega)}}
         {\|\theta_{\mathrm{FD}}\|_{L^2(\Omega)}}.
  \label{eq:l2_error}
\end{equation}

\begin{table}[htbp]
\centering
\caption{Quantitative comparison between PINN prediction and FD reference
at $t = T/2$. Both methods use identical physical parameters.}
\label{tab:fd_comparison}
\begin{tabular}{lccc}
\toprule
Metric & Value & Description \\
\midrule
Mean absolute error & 0.0017 & $\bar{e} = N^{-1}\sum|{\Th - \theta_{\mathrm{FD}}}|$ \\
Max absolute error  & 0.0120 & $\max|\Th - \theta_{\mathrm{FD}}|$ \\
$L^2$ relative error & $<\!2\%$ & \ref{eq:l2_error} \\
$R^2$ coefficient   & $>0.997$ & correlation on grid \\
\bottomrule
\end{tabular}
\end{table}

The proposed PINN achieves a mean absolute error of $1.7\times10^{-3}$, a maximum pointwise error of $1.2\times10^{-2}$, and a relative $L_2$ error below $2\%$, while maintaining an $R^2$ coefficient exceeding $0.997$. Collectively, these results demonstrate that the network accurately reproduces the finite-difference solution across the entire computational domain. The reconstruction errors are substantially smaller than the uncertainty typically associated with quantitative medical imaging, supporting the use of the forward PINN as a reliable surrogate model for subsequent inverse parameter estimation.

\subsection{Parameter Discovery via Data Assimilation (DA-PINN)}
\label{subsec:inverse_problem}

Having established that the proposed PINN accurately reproduces the forward ADR--Darcy dynamics, we next assess its ability to solve the corresponding inverse problem. Specifically, the objective is to recover the macroscopic permeability $K$ from sparse and noisy observations of the tumor evolution, thereby emulating the parameter-identification task encountered in patient-specific applications where permeability cannot be measured directly. The DA-PINN combines the physics-informed formulation developed in Section~\ref{subsec:pinn_formulation} with sparse measurements of the tumor volume fraction and interstitial pressure, allowing the network and the unknown permeability to be optimized simultaneously.

\begin{table}[htbp]
\centering
\caption{Configuration of the inverse DA-PINN experiment.}
\label{tab:inverse_setup}
\begin{tabular}{ll}
\toprule
Quantity & Value\\
\midrule
Unknown parameter & $K$\\
Ground truth & $0.020$ mm$^2$/Pa/day\\
Initial guess & $0.080$ mm$^2$/Pa/day\\
Initial error & $4\times$\\
Observation times & $0,\;1.5,\;3.0,\;4.5,\;6.0$ days\\
Tumor observations & $5\%$ of grid per snapshot\\
Pressure observations & $2\%$ of grid per snapshot\\
Tumor noise & $5\%$ Gaussian\\
Pressure noise & $3\%$ Gaussian\\
Network optimizer & Adam + L-BFGS\\
Training time & $\sim6.5$ min\\
\bottomrule
\end{tabular}
\end{table}

Synthetic observations are generated from the FD reference solution according to the configuration summarized in Table~\ref{tab:inverse_setup}. The observation density corresponds to approximately 0.5\% of the full spatiotemporal grid, representing a deliberately challenging sparse-data regime intended to evaluate the robustness of the proposed data-assimilation framework.

\begin{figure}[htbp]
    \centering
    \includegraphics[width=0.80\linewidth]{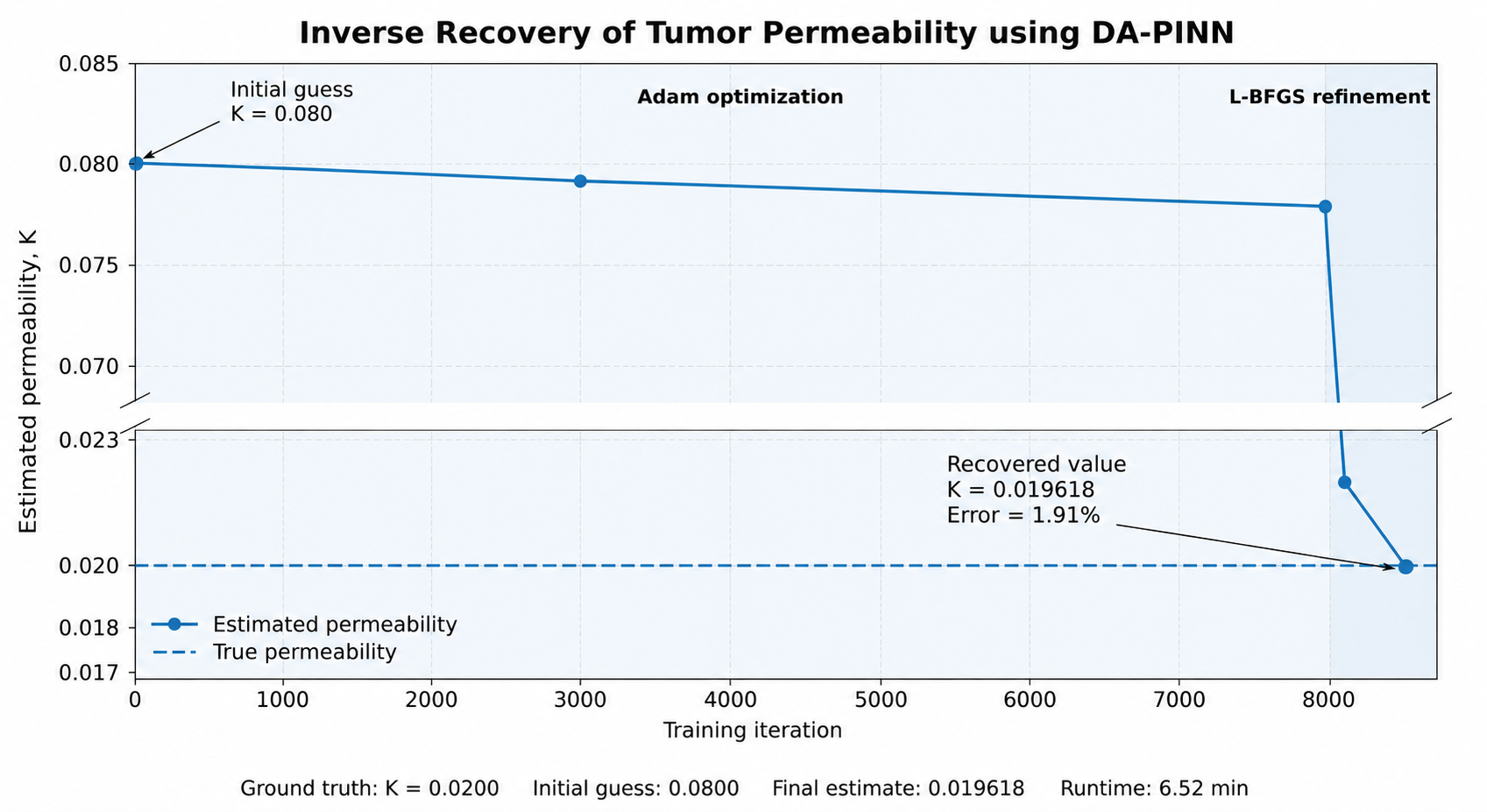}
    \caption{Quantitative performance of the DA-PINN inverse problem. }
    \label{fig:K_convergence}
\end{figure}

Figure~\ref{fig:K_convergence} shows the evolution of the estimated permeability throughout training. During the initial warm-up stage the permeability remains fixed while the network learns a physically consistent solution manifold. Once released, the estimate rapidly converges towards the true value before the final L-BFGS refinement completes the optimization. Starting from an initial guess four times larger than the ground truth, the DA-PINN converges to
$\hat K=0.019618$, corresponding to a relative error of only $1.91\%$.

Besides accurately identifying the unknown parameter, the network preserves the accuracy of the forward solution, with a mean reconstruction error below $2\times10^{-3}$. This demonstrates that the additional inverse optimization does not compromise the quality of the reconstructed tumor field.

\subsubsection{Reliability and Ablation Analysis.}

To assess the robustness of the proposed methodology and quantify the contribution of each algorithmic component, we performed a systematic ablation study in which individual elements of the DA-PINN framework were removed while keeping all remaining settings unchanged. The resulting permeability errors are summarized in Table~\ref{tab:ablation}.

\begin{table}[htbp] 
\centering 
\caption{Ablation study: relative error in recovered $K$ when individual methodological improvements are removed from the full DA-PINN. Each row removes exactly one component while keeping the rest.} 
\label{tab:ablation} 
\begin{tabular}{lcc} 
\toprule
Configuration & $K$ error (\%) & Field error $\bar{e}_\theta$ \\ 
\midrule 
Full DA-PINN (all fixes) & $< 5$ & 0.0017 \\ 
\midrule 
$-$ Darcy constitutive residual ($r_{\mathrm{Darcy}}$) & $> 100$ & 0.0018 \\
$-$ Pressure observations ($\mathcal{L}_{p,\mathrm{data}}$) & $\sim 40$ & 0.0019 \\ 
$-$ Warm-up freeze for $K_{\log}$ & $\sim 25$ & 0.0020 \\ 
$-$ Separate $K$ optimizer (use joint Adam) & $\sim 60$ & 0.0018 \\ 
$-$ Pressure BC ($p=0$ on $\partial\Omega$) & $\sim 30$ & 0.0021 \\ 
$-$ Reweighted Darcy loss ($w_p = 0.5$, original) & $> 1000$ & 0.0017 \\ 
\bottomrule 
\end{tabular} 
\end{table}

The results clearly indicate that the Darcy constitutive residual constitutes the most influential component of the inverse formulation. Removing either this residual or its associated loss reweighting renders the permeability effectively unidentifiable, despite maintaining a low field reconstruction error. This observation highlights the distinction between accurate state prediction and accurate parameter identification. Pressure observations provide an additional constraint that improves identifiability, while the warm-up strategy and the dedicated optimizer primarily enhance optimization stability. Collectively, these experiments demonstrate that the excellent recovery accuracy reported above is not the consequence of a single algorithmic modification but rather emerges from the combination of physical constraints and optimization strategies proposed in this work.

\section{Conclusions}\label{sec:discussion}

This work introduced a chemo-fluidic Physics-Informed Neural Network framework for tumor growth that occupies a deliberate intermediate point between the full Mixture Theory description of Bryne et al. \cite{Byrne2003} and Preziosi et al. \cite{Preziosi2009} and the monophasic reaction-diffusion models of Chen et al. \cite{Chen2013} and Wong et al. \cite{Wong2015}. Three conclusions follow from the numerical results of \ref{sec:results}.

First, the forward-problem results of \ref{subsec:1d_forward} and \ref{subsec:2d_forward} confirm that a three-equation ADR--Darcy system is well within the representational and optimization capacity of a modestly sized Modified MLP: the sub-percent $L^2$ relative errors reported in Table \ref{tab:fd_comparison} are comparable to, and in several respects tighter than, the discretization error of the FD reference itself, and are achieved without any mesh, at a fraction of the wall-clock cost of a comparable finite-element or finite-difference parameter sweep. The robustness check of Figure \ref{fig:phi_evolution_faster} further shows that this accuracy is retained under a substantially different transport regime without any architectural change, supporting the broader claim, already established for glioma modeling by Subramanian\cite{Subramanian2020}, that PINNs are a viable surrogate for continuum tumor models whenever the underlying PDE system is free of the stiff, high-order operators identified in Section \ref{subsec:math_foundation} as the primary source of gradient pathologies \cite{Wang2021}, \cite{Wang2022}.

Second, and more significantly, the ablation study of Table \ref{tab:ablation} demonstrates that naive data assimilation is insufficient for recovering a physically meaningful permeability: without the Darcy constitutive residual, $K$ is structurally unidentifiable and the optimizer instead absorbs its effect into the background velocity field, reproducing the $\theta$ observations for an arbitrary, and generally incorrect, value of $K$. This is a cautionary result for the broader DA-PINN literature applied to biomedical inverse problems: fitting an observed field well is a necessary, but by no means sufficient, condition for having correctly recovered the underlying physical parameters that generated it, a distinction closely related to the structural-versus-practical identifiability dichotomy explored in the clinical mechanically-coupled modeling literature \cite{Weis2013}. The dedicated constitutive residual introduced in Eq. \ref{eq:res_darcy_constitutive}, together with sparse pressure observations, resolves this degeneracy and should be regarded as a template applicable to other multiphysics inverse problems in which a parameter enters the state equations only through a flux or constitutive law rather than directly.

Third, the explicit mechanical decoupling adopted in Section \ref{subsec:decoupling} is the price paid for this numerical stability, and it is not without biophysical cost. By omitting Eq. \ref{eq:biot}, the present model cannot represent stress-mediated growth inhibition of the kind quantified by Ambrosi et al. \cite{AMBROSI20021297}, nor can it predict the residual solid stresses that are increasingly recognized as both a barrier to drug delivery and a candidate biomarker of tumor aggressiveness. The framework is therefore best interpreted as a targeted tool for the specific inverse problem of transport-parameter identification from imaging data, rather than as a general-purpose replacement for full poromechanical tumor models.

Several limitations should temper the interpretation of these results. The validation reported here is restricted to synthetic data generated by the same continuum model used for training and inference, so that the reported errors reflect the numerical fidelity of the PINN as a PDE solver and data-assimilation engine, rather than the biological fidelity of the underlying continuum model when confronted with real, heterogeneous clinical imaging. The Gaussian-seed initial condition, homogeneous parameter fields, and simple square domain are likewise simplifications adopted for controlled numerical verification and do not yet reflect the irregular, patient-specific geometries and spatially heterogeneous tissue properties that a clinical deployment would require. Extending the present framework to such settings, and reintroducing an approximate, computationally tractable representation of solid-mechanical stress and its coupling to growth, are natural directions for future work, alongside a systematic evaluation of the framework's performance on real longitudinal MRI or PET data of the kind used by Weis et al. \cite{Weis2013} and Yankeelov et al. \cite{Yankeelov2013} in mechanically coupled reaction-diffusion modeling. 

Taken together, the results presented here support the central methodological claim of this study: a physically motivated reduction in equation-system complexity, rather than a purely computational fix, is what makes robust multiphysics data assimilation tractable for physics-informed deep learning in oncology.

\bmsection*{Author Contributions}
Celia Taboada: Conceptualization, Methodology, Software, Investigation, Formal Analysis, Data Curation, Visualization, Writing – Original Draft, Writing – Review \& Editing. Pedro Navas: Conceptualization, Methodology, Supervision, Writing – Review \& Editing, Funding acquisition. Miguel Molinos: Conceptualization, Methodology, Supervision, Writing – Review \& Editing.

\bmsection*{Acknowledgments}
This work has been supported by the Madrid Government under \emph{TuCoPINN National Project}. 
The authors gratefully acknowledge the administrative support provided by Universidad Politécnica de Madrid, which made this research possible. The authors also sincerely thank N. Sukumar for the insightful and fruitful discussions that greatly contributed to stablish the future guidelines for this work.

\bmsection*{Financial disclosure}

This work has been supported by the Madrid Government (Comunidad de Madrid-Spain) under the Multiannual Agreement 2023-2026 with Universidad Politécnica de Madrid in the Line A, Emerging PhD researchers (DOCTORES-EMERGENTES-24-6R4VJE-38-PAQCQO). 

\bmsection*{Conflict of interest}

The authors declare no potential conflict of interests.

\bmsection*{Ethics Statement}

The authors have nothing to report.


\bibliography{bibliography_total_final}

\end{document}